\documentstyle[preprint,aps,prl]{revtex}

\newcommand{\be}{\begin{equation}}
\newcommand{\ee}{\end{equation}}
\newcommand{\sst}{\scriptscriptstyle}

\newcommand{\bea}{\begin{eqnarray}}
\newcommand{\eea}{\end{eqnarray}}
\def\r0{r_{\sst 0}}

\begin{document}
\draft
\title{\v{C}erenkov radiation by neutrinos in a supernova core}

\author{\bf{Subhendra Mohanty$^{\dagger} $ and Manoj K. Samal$^{\#} $}}

\address{{\it Theory Group, Physical Research Laboratory, \\
Ahmedabad - 380 009, India }}


\maketitle

\begin{abstract}

{Neutrinos with a magnetic dipole moment propagating in a medium
with a velocity larger than the phase velocity of light emit
photons by the \v{C}erenkov process. The \v{C}erenkov radiation is a
helicity flip process via which a left-handed neutrino in a
supernova core may change into a sterile right-handed one and
free-stream out of the core. Assuming that the luminosity of such
sterile right-handed neutrinos is less than $10^{53}$ ergs/sec
gives an upper bound on the neutrino magnetic dipole moment
$\mu_\nu < 0.2 \times 10^{-13} \mu_B$. This is two orders of magnitude
more stringent than the previously established bounds on $\mu_\nu$
from considerations of supernova cooling rate by right-handed
neutrinos.} 

\end{abstract}
\vskip 1.6cm
\rule{4.3cm}{0.1mm}
\vskip 2mm
$^{\dagger}${\sf E-mail}:{\em mohanty@prl.ernet.in} \hskip 7mm
$^{\#}${\sf E-mail}:{\em mks@prl.ernet.in}
\vskip 1.6cm
\begin{flushleft}
{\sf{PRL-TH-95/10 \\hep-ph/9506385}}
\end{flushleft}
\newpage

The observation of neutrinos \cite{hata} from the supernova
SN1987A has provided a number of constraints on the properties
of neutrinos \cite{nussinov,moha,all}. Most of the mechanisms for the
constraints on the mass and magnetic moment of the neutrinos
depend upon the the helicity flip of a left-handed neutrino into
a sterile right-handed one, which can free-stream out of the core
and hence deplete the energy of the
supernova core within a timescale of $\sim$ 1 sec. Since the
observed time-scale of neutrino emission \cite{hata} is of the order of 1-10
secs, it is expected that the luminosity of the right-handed
neutrinos is less than $10^{53}$ ergs/ sec, which is the
total neutrino luminosity of the supernova. The mechanism
for helicity flip caused by a neutrino magnetic moment that have been
considered so far are (i) helicity flip in an external magnetic
field of the neutron star in the supernova core \cite{nussinov}
and (ii) helicity flip by scattering with charged fermions {\it
i.e.} the processes $\nu_L e^- \rightarrow \nu_R e^-,\: \:\:
\nu_L p \rightarrow \nu_R p$ \cite{moha}. The process (i) leads
to an upper bound $\mu_\nu < 10^{-14} \mu_B$ ($\mu_B = e/ 2
m_e$, the Bohr magneton), but is unreliable since it relies on
a high magnetic field ($\sim 10^{14}$ Gauss) in a supernova core
which has not been observed. The scattering process (ii) leads to
an upper bound $\mu_\nu < (0.2-0.8) \times 10^{-11} \mu_B$
\cite{moha}. 

In this letter we propose a third mechanism for the
neutrino helicity flip which occurs via a \v{C}erenkov radiation
process in the medium of the supernova core. 
In the supernova core the refractive index of photons is
determined by the electric permitivity of the $e^-$, $e^+$
plasma and the paramagnetic susceptibility of the non-relativistic, degenerate 
neutron and proton gas.
We find that \v{C}erenkov emission of a photon from a neutrino is
allowed in the photon frequency range $\omega_p \mu^{1/2}/(\mu
-1)^{1/2} \le \omega \le 2E (\mu^{1/2} -1)/(\mu -1)$, (where
$\omega_p$ is the plasma frequency, $\mu$ is the magnetic
permeability, and  $E$ is the initial neutrino energy). Since the
\v{C}erenkov emission process 
is due to the magnetic dipole operator $\mu_\nu \sigma_{\mu \nu} k_\nu
\epsilon_\mu$, it is a helicity flipping process $\nu_L
\rightarrow \nu_R \gamma$. The helicity flipping is more efficient
in the \v{C}erenkov process because unlike the process (i) there is
no dependence on external magnetic field and unlike (ii) it is a
single vertex process, so the rate is larger than the scattering
rate  $\nu_L e^- \rightarrow \nu_R e^- $ by $({\alpha_{em}}
e^{- \tilde{\mu}_e/ T})^{-1}$, where the exponential factor is due to
the Pauli blocking of the outgoing charged fermion. We compute
the luminosity $Q_{\nu_R}$ of the right-handed neutrinos
produced by the \v{C}erenkov process. The  constraint that
$Q_{\nu_R} < 10^{53}$ ergs/sec (the total observed luminosity)
leads to the bound on the neutrino magnetic moment $\mu_\nu <
0.2 \times 10^{-13} \mu_B$. This is two order of
magnitude improvement over the previously established bound
\cite{moha} owing to the fact that the \v{C}erenkov radiation is
a single vertex process unlike the scattering processes
considered in ref. \cite{moha}. 

The amplitude for the \v{C}erenkov radiation process 
$\nu_L (p) \rightarrow \nu_R (p') \gamma (k)$ is given by 
\be
{\cal M}= \frac{\mu_\nu}{n} \overline{u} (p',s') \sigma^{\mu \nu} k_{\nu} u
(p,s) \epsilon_{\mu} (k, \lambda),
\ee
where $\mu_\nu$ is the magnetic dipole moment of neutrino and
$n$ is the refractive index of the medium. The
transition rate of the \v{C}erenkov process is given by
\be
\Gamma=\frac{1}{2E} \int \frac{d^3 p'}{(2 \pi)^3 2 E'}
\frac{d^3 k}{(2 \pi)^3 2 \omega} (2 \pi)^4 \: \delta^{(4)} (p-p'-k)
|{\cal M}|^2,
\ee
where $p=(E, {\bf p})$, $p'=(E', {\bf p'})$ and $k=(\omega, {\bf
k})$ are the four momenta of the incoming neutrino, outgoing
neutrino and the emitted photon respectively. Using the identity
$$
\int \frac{d^3 p'}{2 E'}= \int d^4 p'  \:\: \theta(E')
\: \delta(p'^2-m_\nu^2),
$$
and integrating over the $\delta$ function in (2) we obtain
\be
\Gamma=\frac{1}{16 \pi} \int \frac{{k}^2 d{k}}{E^2
\omega^2 n}  d(\cos{\theta}) \: \delta (\frac{2 \omega E-k^2}{2 |{\bf
k}||{\bf p}|}- \cos{\theta}) |{\cal M}|^2,
\ee
where ${\theta}$ is the angle between the emitted photon and
the incoming neutrino. 

In a medium with the refractive index $n (= |{\bf k}|/\omega)$,
the $\delta$ function in (3) constrains $\cos{\theta}$ to
have the value
\be
\cos{\theta}=\frac{1}{n v}[1 + \frac{(n^2 -1) \omega}{2 E}],
\ee
where $v= |{\bf p}|/E$ is the particle velocity and $k^2=- (n^2
-1) \omega^2$. It is clear that the
kinematically allowed region for the \v{C}erenkov process is where
$|\cos{\theta}|$ given by (4) is $ \le 1$. 

Evaluating $|{\cal M}|^2$ from (1) and substituting it in (3)
and performing the integral over $\delta$ function and using (4) for
$\cos{\theta}$, we have the expression for transition rate for
the \v{C}erenkov process \cite{bell,ginz,grimus}
\be
\Gamma=\frac{\mu_\nu^2}{16 \pi E^2} \int_{\omega_1}^{\omega_2} d 
\omega \frac{(n^2-1)^2}{n^2} [ \{4E^2 + 4 m_\nu^2
\frac{n^2}{(n^2-1)}\} \omega^2- 4 E \omega^3- (n^2-1) \omega^4 ],   
\ee
where the limits of the integral give the range of frequency
allowed for the {\v C}erenkov photon and the refractive index
$n$ is, in general, a function of $\omega$.

The refractive index of photons in a medium can be expressed as
$ n^2= \epsilon \mu$, where $\epsilon$ and $\mu$ are the
electric permitivity and 
magnetic permiability of the medium. In the supernova core, 
the medium is a plasma consisting of degenerate electrons,
protons and neutrons at temperature $T \approx 60$
MeV \cite {moha,turner}. The permitivity $\epsilon$ is given by
\be
\epsilon= (1- \frac{\omega_p^2}{\omega^2}),
\ee
where $\omega_p=(4 \alpha_{em} /3 \pi)^{1/2} \tilde{\mu}_e $
\cite{salpeter} is the plasma frequency that is determined by 
the chemical potential of the electrons $\tilde{\mu}_e$.  
In a non-magnetic plasma where magnetic permitivity $\mu=1$, the
refractive index $n= (1-\omega_p^2/\omega^2)^{1/2} < 1$ and the
\v{C}ernkov process is therefore forbidden. In the
supernova core, however there is a large density of
non-relativistic neutrons and protons which
contribute to the paramagnetic susceptibility $\chi$, ( related
to the magnetic permiability $\mu= 1+ 4 \pi \chi$) through their
magnetic dipole moments.  

Treating the neutrons and protons in the supernova core as degenerate
Fermi gas, the magnetic susceptibility becomes independent of
temperature and is given as a function of the photon wavelenght $k$
as  \cite{landau,qtss}
\be 
\chi_i(k)= \frac{1}{2 \pi^2}(2 m_i)^{3/2} {\mu_i}^2 {\tilde{\mu}_i}^{1/2}
  ({1\over 2})(1+ {4{k_{fi}}^2 -k^2 \over 4 k_{fi} ~ k }~ln |{2k_{fi} +k
\over 2k_{fi} - k}| ~~ )  ,
\label{param}
\ee
where $m_i$ is the mass, $\mu_i$ is the magnetic moment, $\tilde{\mu}_i$
is the chemical potential  and 
$k_{fi} = (2 m_i \tilde{ \mu_i})^{1/2}$ is the fermi momentum 
 of the $i \:th$  fermion
species. In the supernova core the photon wavelenghts
$|k|\sim T << k_{fi} $ as $T\sim 60 Mev $ and the fermi momentum
$k_{fn} \sim 894 MeV $ for neutrons ( with
chemical potential $\tilde \mu_n \sim 400 MeV$ )
 and $k_{fp} \sim 748
MeV $ for protons ( with $\tilde \mu_p \sim 280 MeV$) \cite{moha,all}. The
wavenumber dependence of the paramagnetic susceptibility (\ref{param}) 
can be expanded as a series in the small parameter $(k/k_f)$ as
\be 
\chi_i(k)= \frac{1}{2 \pi^2}(2 m_i)^{3/2} {\mu_i}^2 {\tilde{\mu}_i}^{1/2}
  (1-  ({k \over 2 k_{fi}})^2 + \cdots   ) ,
\label{param2}
\ee
The contribution of the wavenumber dependent terms is about $10^{-3}$
smaller compared to the leading order term and will  be
neglected in 
the present analysis. 
 Only the non-relativistic fermions
(neutrons and protons) contribute to the dipole susceptibility
\cite{landau}.
If the photon frequency is close to the nucleon mass,
such that the photon could produce significant numbers of nucleon
pairs during propagation  then the expression 
(7) for the diamagnetic susceptibity will  no longer be valid. Therefore our
 conclusions are subject to the assumption that the photon freqencies
(which we assume to be close to the temperature $T \sim 60 MeV$) are small
compared to the nucleon masses of $\sim 1 GeV$. 
In our treatment we do not consider the corrections to (7) due to 
 pair production
\cite{qtss} at frequencies close to the nucleon masses. 
 Taking the chemical potentials $\tilde{\mu_p}
\approx 280$ MeV, $\tilde{\mu_n} \approx 400 $ MeV \cite{moha,all}
for the protons and neutrons respectively, the susceptibility 
 $\chi=\chi_p +\chi_n = 0.15$ and the magnetic permiability
turns out to be $\mu=2.88$. 
The electrons being relativistic $(m <<
T)$, do not contribute to the magnetic susceptibility. The
electrons with chemical potential $\tilde{\mu_e} \approx 280$
MeV and plasma frequency $\omega_p$ contribute to the refractive
index via the electric permitivity $\epsilon$.
Consequently, the refractive index of the non-relativistic
degenerate neutrons and protons and relativistic electrons in a
supernova core is given by 
\be
n(\omega)= (\mu \epsilon)^{1/2}= [1+ 4 \pi (\chi_n + \chi_p)]^{1/2} (1 -
\frac{\omega_p^2}{\omega^2})^{1/2}.
\ee

By combining the kinematic constraint (4)
with the above expression for the refractive index (9), we find that
\v{C}erenkov radiation by neutrinos is kinematically allowed for
the range of the frequency $\omega$ given by 
\be
\frac{\omega_p \mu^{1/2}}{(\mu -1)^{1/2}} \le \omega \le
\frac{2E (\mu^{1/2} -1)}{(\mu -1)}, 
\ee
where we have taken $|p|= E (1-m^2/E^2)^{1/2} \simeq E$ since we
are dealing with extremely relativistic neutrinos with $m^2/E^2
< 10^{-12}$.

Keeping terms upto second order in $\omega$ in the expression
for $n(\omega)$, and neglecting the second term (since $m_\nu \sim 1$ eV
and $\omega \sim 60 $ MeV) in the expression for $\Gamma$ given
in (5), the transition rate for the {\v C}erenkov process
in the supernova core is evaluated to be
\begin{eqnarray}
\Gamma&=&\frac{\mu_\nu^2}{16 \pi E^2} \frac{(\mu -1)}{\mu}
\int_{\omega_1}^{\omega_2} d \omega [4 E (\omega-E) \{(1-\mu)
\omega^2 + \omega_p^2 (\mu +1) \} \nonumber\\ 
&&+ (\mu-1) \omega^2 \{ (1- \mu)
\omega^2 + \omega_p^2 ( 2 \mu+1) \}] \nonumber\\ 
&=&\frac{\mu_\nu^2 E}{6 \pi \mu^2} [\frac{2}{5} E^2
\frac{(\mu^{1/2} -1)^2 (4 \mu^{3/2} + 16 \mu -15)}{(1+
\mu^{1/2})^2} + \omega_p^2 \mu (1- \mu^{3/2})].   
\end{eqnarray}

The \v{C}erenkov process $\nu_L (p) \rightarrow \nu_R (p') \gamma (k)$
changes the $\nu_L$'s to sterile $\nu_R$'s which can free
Stream out of the supernova core. The luminosity of the sterile
$\nu_R$'s is the product of the energy taken by each
right-handed particle {\it i.e.} $(E-\omega)$ and the total number of
right-handed particles produced per unit volume as given by
(11), multiplied with the volume of the supernova core and is
found to be
\begin{eqnarray} 
Q_{\nu_R}&= &\frac{ 3 V \mu_\nu^2}{16 \pi} \frac{(\mu -1)}{\mu}
\int_{0}^{\infty} \frac{dE}{E^2}
[f_\nu(E)-f_{\overline{\nu}}(E)] E^2 \int_{\omega_1}^{\omega_2} d
\omega (E-\omega) \times \nonumber\\ 
&& [4 E (\omega-E) \{(1-\mu) \omega^2 + \omega_p^2 (\mu +1) \} +
(\mu-1) \omega^2 \{ (1- \mu) \omega^2 + \omega_p^2 (2 \mu +1) \}],
\end{eqnarray}
where $f_\nu(E) = [e^{(E-\tilde{\mu}_\nu)/T}+1]^{-1}$ and 
$f_{\overline{\nu}}(E) = [e^{(E+\tilde{\mu}_\nu)/T}+1]^{-1}$ are
the statistical distribution function of the $\nu_L$ and
$\overline{\nu_L}$ in the supernova core, $\tilde{\mu_\nu}$ is
the chemical potential of the neutrino and the factor of $3$ is due
to the contributions from all three neutrino flavours as the
cooling proceeds through the emission of $\nu \overline{\nu}$
pairs of all flavours, created in thermal equilibrium 
\cite{stanev}. Performing the
integrals over $E$, the luminosity of right handed
neutrinos is obtained as
\begin{eqnarray}
Q_{\nu_R}&=& \frac{V \mu_\nu^2 \tilde{\mu_{\nu}}
(\mu^{1/2}-1)}{210 \pi \mu (1+\mu^{1/2})^3} 
[16 {\tilde{\mu_{\nu}}}^2  (3 {\tilde{\mu_{\nu}}}^4 + 21 \pi^2
T^2 {\tilde{\mu_{\nu}}}^2 + 49 \pi^4 T^4) \mu (\mu^{1/2}-1)
\nonumber\\  
&& - 7 \omega_p^2 (3 {\tilde{\mu_{\nu}}}^4 + 10 \pi^2 T^2
{\tilde{\mu_{\nu}}}^2 + 7 \pi^4 T^4) (1+4 \mu^{1/2} +7 \mu +10
\mu^{3/2} + 8 \mu^2 + 2 \mu^{5/2})].   
\end{eqnarray}
We take the volume $V \approx 4 \times 10 ^{18}$ cm$^3$,
$\tilde{\mu_\nu} \approx 160$ MeV, $T \approx 60$ MeV \cite{moha}
for the supernova core parameters within 1 second after collapse.
Using these numbers we obtain the luminosity to be
\be
Q_{\nu_R}= 0.98 \times 10^{53} \:\:\: \mu_\nu^2  \:\:\: GeV^4, \label{htlos}
\ee
in terms of the magnetic moment of neutrino $\mu_\nu$. 
Assuming that the entire energy of the core collapse is not
carried out by the right handed sterile 
neutrinos, {\it i.e.} $Q_{\nu_R} < 10^{53}$ ergs/ sec, we have
from (\ref{htlos}) the upper bound on the neutrino magnetic
dipole moment given by
\be
\mu_\nu < 0.2 \times 10^{-13} \mu_B. \label{upbnd}
\ee
Varying the core temperature of the supernova in the range $30-
70$ MeV, the upper bound (\ref{upbnd}) is seen to fall in the
range $ (0.59-0.15)\times 10^{-13} \mu_B$ respectively.
The upper bound on neutrino magnetic moment given in (\ref{upbnd}) is two
orders of magnitude better than the previously 
established \cite{moha} upper bound from the $\nu_R$ luminosity
of supernova. The process for generating $\nu_R$ in the
supernova core considered in ref. \cite{moha} is via the
helicity flip scattering $\nu_L e^- \rightarrow e^- \nu_R$ and
$\nu_L p^- \rightarrow p^- \nu_R$ etc. This process has an extra
electromagnetic vertex and a Pauli blocking factor for the
outgoing charged fermion compared to the process that we have
considered and is suppressed compared to the process considered
here by the factor ${\alpha_{em}}\: e^{- \tilde{\mu_e}/T}$.
That accounts for the more stringent bound we 
have  compared to ref. \cite{moha}.

\centerline{\bf \large Acknowledgement}

We thank Walter Grimus for useful correspondence and
for drawing our attention to ref \cite{grimus} where the
possibility of detection of transition \v{C}erenkov radiation by
neutrinos is studied.

\end{document}